\documentclass[aps,prl,twocolumn,superscriptaddress,showpacs]{revtex4-1}

\usepackage{bm}
\usepackage{color}
\usepackage{amsmath}
\usepackage[dvips]{graphicx}
\usepackage{textcomp}

\begin{document}


\title{Novel Phase Transitions in the Breathing Pyrochlore Lattice:\\
$^{7}$Li-NMR on LiInCr$_4$O$_8$ and LiGaCr$_4$O$_8$}


\author{Yu Tanaka}
\affiliation{Institute for Solid State Physics, University of Tokyo, Kashiwa, Chiba 277-8581, Japan}
\author{Makoto Yoshida}
\affiliation{Institute for Solid State Physics, University of Tokyo, Kashiwa, Chiba 277-8581, Japan}
\author{Masashi Takigawa}
\affiliation{Institute for Solid State Physics, University of Tokyo, Kashiwa, Chiba 277-8581, Japan}
\author{Yoshihiko Okamoto}
\altaffiliation[Present address: ]{Department of Applied Physics, Nagoya University, Furo-cho, Chikusa-ku, Nagoya 464-8603, Japan}
\affiliation{Institute for Solid State Physics, University of Tokyo, Kashiwa, Chiba 277-8581, Japan}
\author{Zenji Hiroi}
\affiliation{Institute for Solid State Physics, University of Tokyo, Kashiwa, Chiba 277-8581, Japan}


\date{\today}

\begin{abstract}
We report $^{7}$Li-NMR studies on LiInCr$_4$O$_8$ and LiGaCr$_4$O$_8$, in which Cr$^{3+}$ 
ions  with spin 3/2 form a breathing pyrochlore lattice, a network of tetrahedra with alternating sizes.
In LiInCr$_4$O$_8$ with large alternation, the nuclear relaxation rate 1/$T_1$ shows 
an activated temperature ($T$) dependence down to 18~K, indicating a singlet ground state 
with a spin gap. This behavior, however, is disrupted by an antiferromagnetic (AF) transition at 
13~K, which is preceded by another, most likely structural, transition at 16~K.   In contrast, 
LiGaCr$_4$O$_8$ with small alternation shows no spin gap but exhibits a first-order 
AF transition over a distributed $T$-range 13--20~K. Nevertheless,
1/$T_1$ of the paramagnetic phase diverges toward 13~K, indicating proximity to a 
second-order transition. The results indicate that LiGaCr$_4$O$_8$ is located in the 
vicinity of a tricritical point in the phase diagram.
\end{abstract}


\maketitle


 Geometrically frustrated spin systems have been extensively studied from both 
experimental and theoretical aspects because frustrating interactions prevent conventional 
magnetic order and may lead to an exotic ground state. The pyrochlore lattice, a three dimensional 
network of corner-sharing tetrahedra, is a well-known example  of frustrated geometry. 
The ground state of Heisenberg spins on this lattice with an
antiferromagnetic (AF) interaction between nearest neighbors is theoretically predicted to be a 
spin liquid without long-range magnetic order~\cite{Moessner,Canals_lett,Canals_B,Tsunetsugu}.  
In real materials, however, an ordered state can be stabilized by various kinds of perturbations~\cite{Bramwell,Yamashita,Tchernyshyov,Bergman,Palmer,Elhajal,Reimers,Bellier-Castella}.
Therefore, definite observation of a spin liquid state has not been reported yet.
Cr spinel oxides {\textit A}Cr$_2$O$_4$ ({\textit A} = Mg, Zn, Cd, Hg) where Cr$^{3+}$ ions with
spin 3/2 form a pyrochlore lattice, have been investigated as good model materials~\cite{Takagi}. 
Although magnetic orders in these materials are largely depressed by frustration,
simultaneous AF and structural first-order transitions occur at low
temperatures~\cite{Ortega,Lee_Gasparovic,Lee_Broholm,Matsuda}.  The spin-Jahn-Teller effect due to
spin-lattice coupling has been proposed as the mechanism for these 
transitions~\cite{Yamashita,Tchernyshyov,Sushkov}. 
The spin-lattice coupling is considered to play important roles also for various phases in high 
magnetic fields including the magnetization plateaus~\cite{Ueda_Katori,Miyata,Ueda_Mitamura,Penc}.

\begin{figure}[b]
	\includegraphics[width=0.9\linewidth]{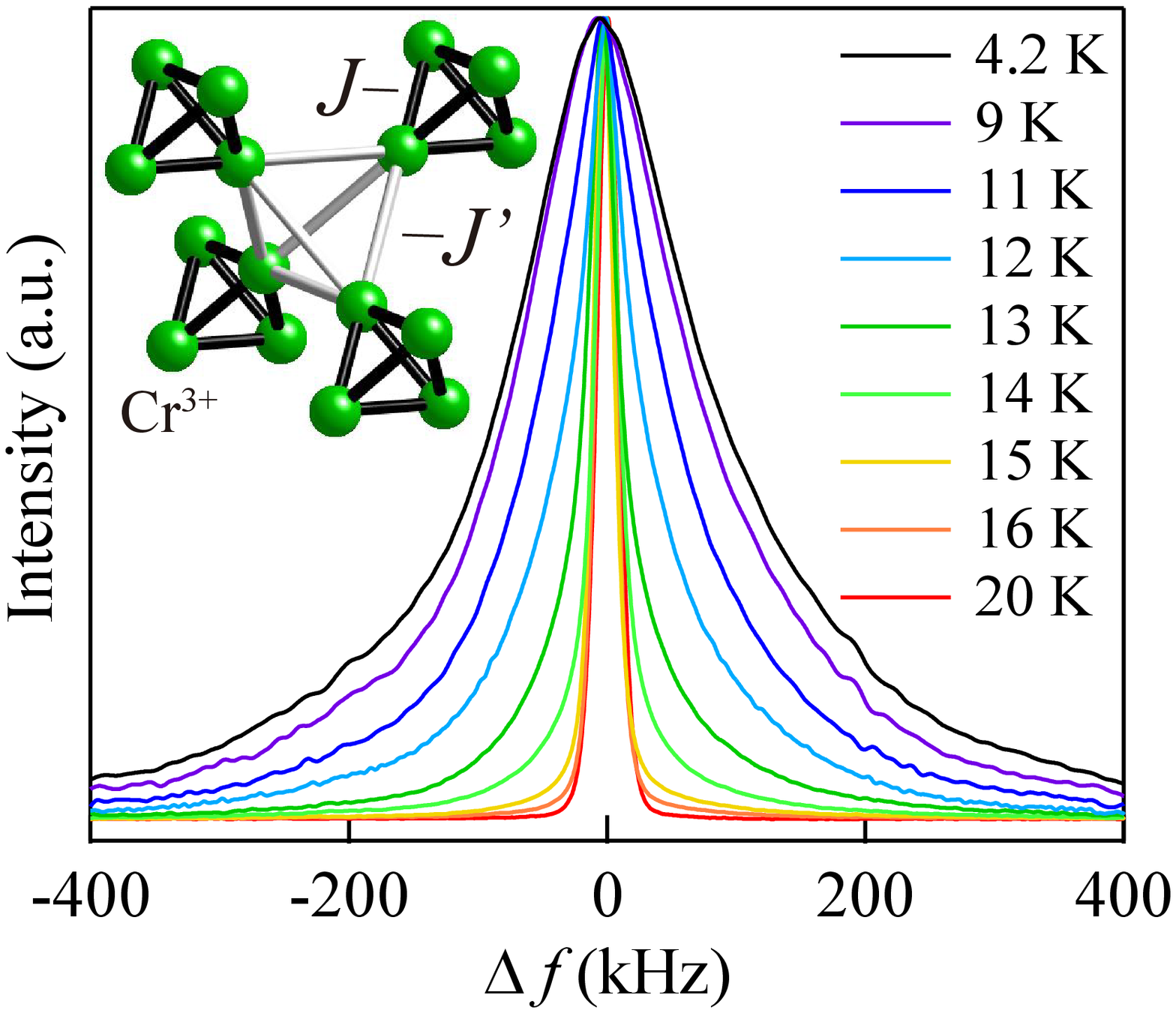}%
 \caption{(Color online) Temperature dependence of the NMR spectra for LiInCr$_4$O$_8$ obtained at 2~T. 
The vertical scale is normalized by the peak intensity for each spectrum. The origin of the 
horizontal axis $\Delta f = 0$ corresponds to the center of gravity of the spectrum. 
The inset shows the structure of a breathing pyrochlore lattice.}
 \label{FIG1}
 \end{figure}
    
Recently, Okamoto \textit{et al.} reported magnetic properties of a new type of Cr spinel oxides 
LiInCr$_4$O$_8$ and LiGaCr$_4$O$_8$~\cite{Okamoto}. The Li$^{+}$ and In$^{3+}$/Ga$^{3+}$ ions 
in these materials alternately occupy the $A$-site of $A$Cr$_2$O$_4$~\cite{Joubert}. This results in
alternation of the nearest Cr-Cr bond length in all directions (see the inset of Fig.~\ref{FIG1}). The Cr spins,
therefore, form a breathing pyrochlore lattice with alternation of large and small tetrahedra and two distinct
exchange couplings $J$ and $J^{\prime}$ ($J > J^{\prime}$). The breathing factor $B_{\mathrm f}$ defined
as $B_{\mathrm f}=J^{\prime}/J$ quantifies the degree of frustration, which is estimated to be 0.1 for 
LiInCr$_4$O$_8$ and 0.6 for LiGaCr$_4$O$_8$~\cite{Okamoto}. The two limiting values 
$B_{\mathrm f}$ = 0 and 1 correspond to the cases of trivial isolated tetrahedra 
and strongly frustrated uniform pyrochlore lattice, respectively.
Therefore, spin systems on the 
breathing pyrochlore lattice with various values of $B_{\mathrm f}$ provide unique opportunity 
to tune the frustration and possibility to discover novel quantum phases. 
The magnetic susceptibility $\chi$ of LiInCr$_4$O$_8$ shows an activated temperature dependence, 
which can be fit reasonably well to the result of an isolated tetrahedron with a spin gap of
57~K~\cite{Okamoto}. In contrast, $\chi(T)$ of LiGaCr$_4$O$_8$ is similar to that of 
ZnCr$_2$O$_4$~\cite{Okamoto}. The heat capacity divided by temperature $C_{p}/T$ 
exhibits a sharp peak at 15.9~K for LiInCr$_4$O$_8$ and at 13.8~K for LiGaCr$_4$O$_8$, indicating a phase transition in both compounds~\cite{Okamoto}. 
   
We have performed $^{7}$Li-NMR measurements on LiInCr$_4$O$_8$ and LiGaCr$_4$O$_8$
to get microscopic understanding of these phase transitions. 
The nuclear spin-lattice relaxation rate 1/$T_1$ in LiInCr$_4$O$_8$ exhibits an activated temperature 
dependence down to 18~K, suggesting a singlet ground state with an excitation gap.
This behavior, however, is disrupted by successive phase transitions.
On the other hand, LiGaCr$_4$O$_8$ shows no sign of a spin 
gap and undergoes a first-order AF transition. 
However, 1/$T_1$ of the paramagnetic phase 
exhibits a critical divergence toward 12.8~K as if there were a second-order magnetic transition. 
Our results indicate that LiGaCr$_4$O$_8$ is located in the vicinity of a tricritical 
point in the phase diagram. 
This finding opens a new route to study spin-lattice coupling on a pyrochlore lattice. 


The powder samples of LiInCr$_4$O$_8$ and LiGaCr$_4$O$_8$ were synthesized by the solid state reaction
method~\cite{Okamoto}. 
All the  $^{7}$Li-NMR measurements were done in a magnetic field of 2~T. 
In both compounds, the NMR spectra consist of a sharp single line in the paramagnetic state with no 
quadrupole structure, consistent with the cubic local symmetry $\overline{4}3m$ 
at the $^{7}$Li site. 
The hyperfine coupling constant was estimated to be 0.09~T/$\mu_{\mathrm{B}}$ for LiInCr$_4$O$_8$ from the magnetic shift and $\chi$ measured
above 18~K and 0.1~T/$\mu_{\mathrm B}$ for LiGaCr$_4$O$_8$ above 80~K.
The NMR spectra were obtained by Fourier transforming the spin-echo signal. 
When the spectra became broad in the AF state, the whole spectra were constructed  
by summing the Fourier transformed spin-echo signals obtained at equally spaced frequencies. 
1/$T_1$ was determined by fitting the recovery curves of the spin-echo 
intensity $I(t)$ as a function of the time $t$ after an inversion pulse to the stretched-exponential function 
	\begin{equation}\label{eq:streach}
  	I(t)=I_{\mathrm{eq}}-I_0  \exp \left[-\left(\frac{t}{T_1}\right)^{\beta}\right]~~~ (0<\beta \leq1),
	\end{equation}
where $I_{\mathrm{eq}}$ is the intensity at the thermal equilibrium and $\beta$ is the stretch exponent that 
provides a measure of inhomogeneous distribution of 1/$T_1$. The case of homogenous relaxation
corresponds to $\beta$ = 1. For broad spectra, 1/$T_1$ was measured at the spectral center.

\begin{figure}[b]
	\includegraphics[width=0.95\linewidth]{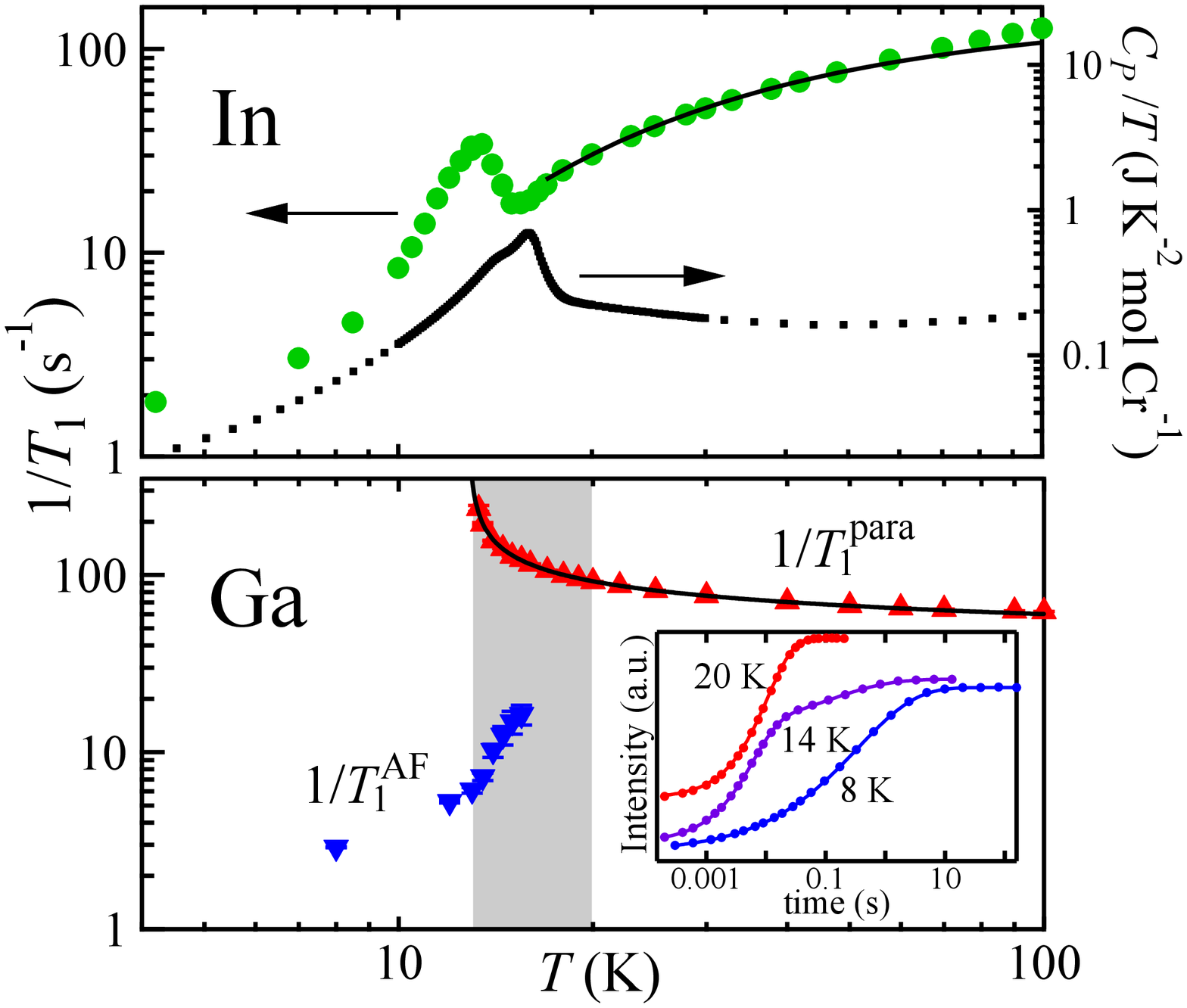}%
\caption {(Color online) Upper panel: Temperature dependences of 1/$T_1$ measured at 2~T 
for LiInCr$_4$O$_8$ with a fit to the activation law and the $C_{p}/T$ data at zero field
taken from \cite{Okamoto}. Lower panel: Temperature dependences of $1/T_1^{\mathrm{para}}$ and 
$1/T_1^{\mathrm{AF}}$ measured at 2~T for LiGaCr$_4$O$_8$. 
The shaded area represents the coexistence region determined by Fig.~\ref{FIG3}(c).
The solid line is the fit to Eq.~(\ref{eq:Moriya}).
The inset shows the recovery curves at different temperatures.}
 \label{FIG2} 
 \end{figure}

We first discuss the results on LiInCr$_4$O$_8$. Figure~\ref{FIG1} shows the temperature dependence of the NMR spectra. 
The spectrum at 20~K has a sharp line of the paramagnetic phase. With decreasing temperature, the spectrum broadens continuously below 14~K, indicating an AF transition.
The upper panel of Fig.~\ref{FIG2} shows the temperature dependence of 1/$T_1$ for LiInCr$_4$O$_8$ 
compared with the $C_{p}/T$ data reported in \cite{Okamoto}.
The recovery curves can be fit well to the single exponential function described by Eq.~(\ref{eq:streach}) 
with $\beta$ = 1 above 16~K, indicating homogeneous relaxation. 
Below 16~K, $\beta$ gradually decreases and reaches 0.5 at 4.2~K.
In the temperature range 18--48~K, 1/$T_1$ can be fit well to the activation law, $1/T_1 
\propto \exp \left(-\varDelta/T\right)$, with the energy gap $\varDelta$ = 31~K 
(the solid line). This result is
consistent with the singlet formation expected for a breathing pyrochlore lattice with small $B_{\mathrm f}$.
The value $\varDelta$ = 31~K is somewhat smaller than, but comparable to, $\varDelta$ = 56.8~K estimated 
by fitting $\chi(T)$ to the model of an isolated tetrahedron~\cite{Okamoto}.
However, $1/T_1$ increases suddenly below 16~K and shows a sharp peak at 13~K, 
indicating an AF transition with critical slowing down  with $T_{\mathrm{N}} = 13$~K. 

This value of $T_{\mathrm{N}}$ is clearly different from the peak 
temperature $T_{\textrm P}$ = 15.9~K in $C_{p}/T$. 
In general, NMR measurements are very sensitive to a magnetic phase transition, 
but are not sensitive to a structural one if the sample is polycrystalline. 
Therefore, it is most likely that the peak in $C_{p}/T$ corresponds to a structural transition, 
not a magnetic one, which changes the symmetry of the lattice and thereby releases the frustration. 
This then enables a second-order AF transition at a slightly lower temperature. 
A preliminary neutron scattering experiment indicates symmetry lowering from cubic to probably orthorhombic 
in the low temperature phase~\cite{Nilsen}, supporting our scenario. 
The small shoulder in $C_{p}/T$ around 14~K could be attributed to this AF transition.
We also discuss this two-step transition from the viewpoint of the spectral line width in the Supplemental Material~\cite{SM}.

\begin{figure}[t]
	\includegraphics[width=0.95\linewidth]{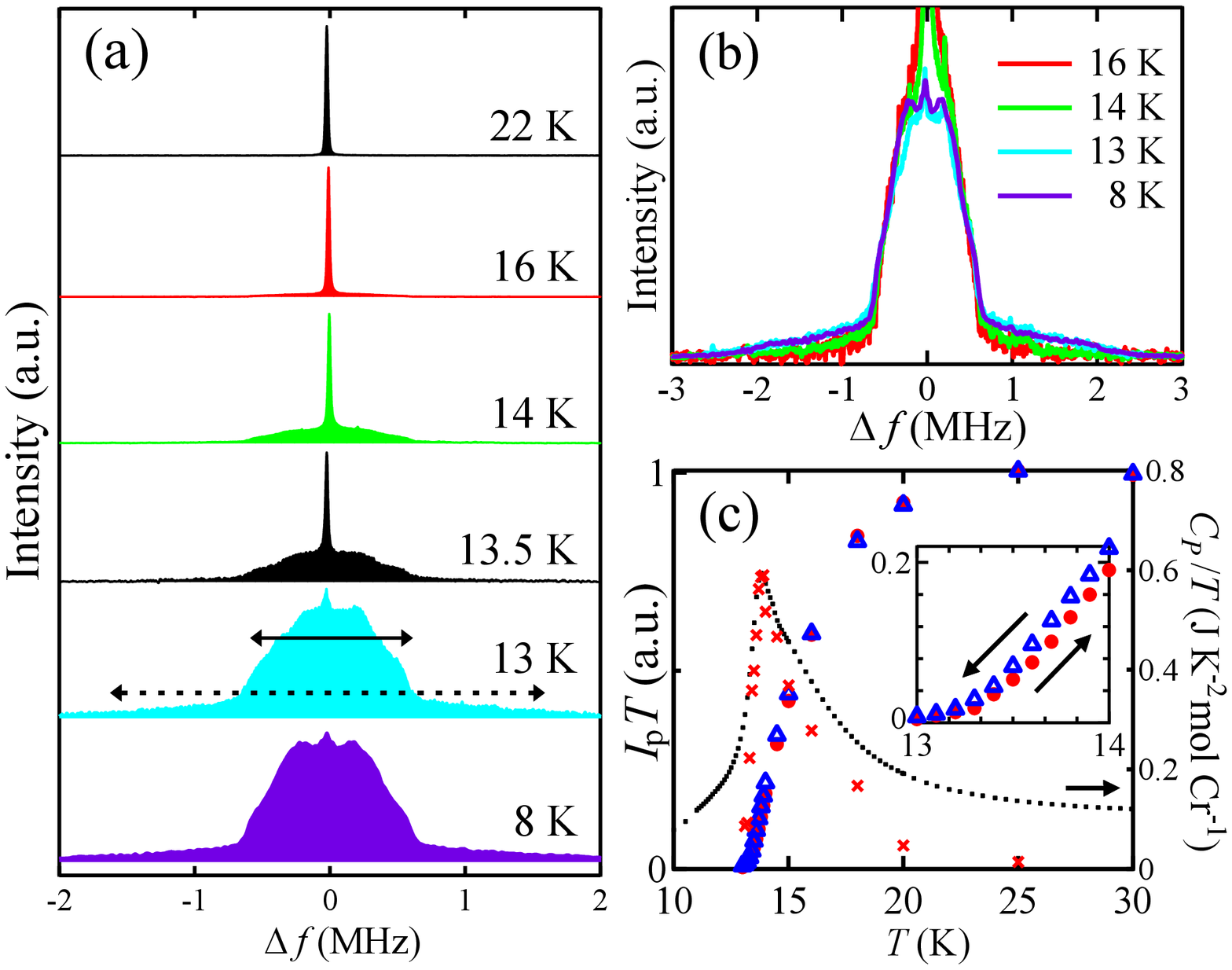}%
 \caption{
 (Color online) (a) Temperature dependence of the NMR spectra for LiGaCr$_4$O$_8$ obtained at 2~T.  
 The vertical scale is normalized by the peak intensity. 
 The origin of the horizontal axis $\Delta f = 0$ corresponds to the center of gravity.
 (b) The AF components of the spectra below 16~K. The vertical scale is normalized by the top of the broad line for each 
spectrum, ignoring the sharp paramagnetic component. 
(c)  The solid red circles (open blue triangles) show the intensity of the paramagnetic line 
multiplied by temperature $I_{\mathrm{p}} T$, which represents the 
paramagnetic volume fraction, measured with heating (cooling).  The same data are expanded in the inset. 
The black dots show the $C_{p}/T$ data at zero field reported in~\cite{Okamoto}. 
The red crosses show $d (I_{\mathrm{p}} T)/dT$ obtained from the central difference 
of the discrete $I_{\mathrm{p}} T$ data.
}
 \label{FIG3}  
 \end{figure}

We now turn to the results on LiGaCr$_4$O$_8$. Figure~\ref{FIG3}(a) shows the temperature dependence of the NMR spectra. 
A sharp line of the paramagnetic phase is observed above 22~K with the line width comparable to that of LiInCr$_4$O$_8$. In the temperature range 13.5--16~K, the spectrum consists of a 
sharp paramagnetic line and a broad line originating from an AF phase, indicating coexistence of 
two phases.  The paramagnetic component completely disappears below 13~K. Although the spectra 
at 13~K and 8~K show a small peak near the center, it is likely from a minor non-magnetic impurity phase.  
The broad spectrum from the AF phase further consists of two components: a relatively narrow line whose
FWHM is about 1~MHz and a much wider one (the solid and dotted arrows in Fig.~\ref{FIG3}(a)). 

Generally, the line width in an AF phase is proportional to the magnitude of the ordered moments. 
Therefore, temperature dependence of the line width provides information about how the AF moments 
develop with temperature. For instance, the gradual line broadening observed in LiInCr$_4$O$_8$ below 14~K 
(Fig.~\ref{FIG1}) indicates a continuous growth of the AF moments associated with a second-order 
transition. Figure~\ref{FIG3}(b) shows the broad components of the spectra for LiGaCr$_4$O$_8$ below 16~K.
We can clearly see that all the spectra show 
nearly identical line shape except for small difference in the intensity of the wide component. 
This means that the AF moments develop discontinuously and the magnitude of 
the ordered moments is independent of temperature once the transition has occurred, even though the
macroscopic state is inhomogeneous mixing of the two phases. Therefore, we conclude that the AF 
transition in LiGaCr$_4$O$_8$ is first-order. 
 
The volume fraction of the paramagnetic phase changes gradually over a rather wide
temperature range as indicated in Fig.~\ref{FIG3}(c). Here, the integrated intensity of the sharp paramagnetic
line multiplied by temperature $I_{\mathrm{p}} T$, which represents the volume fraction of the 
paramagnetic phase, is plotted against temperature  (red solid circles and blue open triangles).  
A small hysteresis is observed between the heating and cooling processes, supporting the first-order 
nature of the AF transition.  The gradual change of the paramagnetic volume fraction for 13--20~K 
 indicates distribution of the AF transition temperature, consistent with the 
macroscopic coexistence of the two phases. The distribution function of the transition 
temperature obtained by differentiating $I_{\mathrm{p}} T$ with $T$ (red crosses) shows 
a peak near 14~K similar to $C_{p}/T$ reported in \cite{Okamoto} (black dots). 

Next we discuss the  1/$T_1$ data for LiGaCr$_4$O$_8$.
The recovery curves $I(t)$ measured at different temperatures are displayed in the inset of the lower panel of Fig.~\ref{FIG2}. 
The single-exponential function described by Eq.~(\ref{eq:streach}) with $\beta  = 1$ gives 
a good fit above 20~K, where only the paramagnetic component is present. 
On the other hand, $I(t)$ in the AF phase below 13~K can be fit to the stretched-exponential 
function with $\beta \sim$ 0.4. In the intermediate temperature region 13.3--16~K, $I(t)$ cannot be fit by 
Eq.~(\ref{eq:streach}) with any $\beta$. Since the paramagnetic and AF phases coexist in this
temperature region, we determined 1/$T_1$ for the
paramagnetic phase (1/$T_1^{\mathrm{para}}$) and the AF phase (1/$T_1^{\mathrm{AF}}$) separately 
by using a sum of the single-exponential and the stretched-exponential functions,
\begin{equation}\label{eq:single&streach}
	I(t)=I_{\mathrm{eq}}-I_{\mathrm{para}}  \exp \left(-\frac{t}{T_1^{\mathrm{para}}} \right) -I_{\mathrm{AF}}  
\exp \left[-\left(\frac{t}{T_1^{\mathrm{AF}}}\right)^{\beta}  \right].
\end{equation}
This equation gives a good fit to the recovery curve in the coexistence region 
as displayed in the inset of the lower panel of Fig.~\ref{FIG2}.

The lower panel of Fig.~\ref{FIG2} shows the temperature dependences of  $1/T_1^{\mathrm{para}}$ 
and $1/T_1^{\mathrm{AF}}$ for LiGaCr$_4$O$_8$. 
1/$T_1^{\mathrm{para}}$ increases slightly with decreasing 
temperature at high temperatures without any sign of a spin gap in contrast to the result in LiInCr$_4$O$_8$.
With further decreasing temperature, 1/$T_1^{\mathrm{para}}$ shows a divergence near 13~K, which indicates
critical slowing down of magnetic fluctuations. Such a divergence is usually associated with a 
second-order magnetic transition and seemingly contradicts our conclusion for the first-order transition
based on the temperature dependence of the NMR spectra. 
The temperature dependence of 
1/$T_1^{\mathrm{para}}$ can be fit to the function
      \begin{equation}\label{eq:Moriya}
  	\frac{1}{T_1^{\mathrm{para}}}=\frac{A}{\sqrt{T-T_{\mathrm{N}}}}+{\mathrm{const}}, 
      \end{equation}
derived for unfrustrated antiferromagnets near the transition temperature 
$T_{\mathrm{N}}$~\cite{Moriya}. We obtain $T_{\mathrm{N}}$ = 12.8~K as shown by the solid line in the lower 
panel of Fig.~\ref{FIG2}. 

In a pure spin system on a breathing pyrochlore with the limiting value $B_{\mathrm f}$ = 0 or 1, 
a second-order AF transition is not expected. 
It is an interesting result that LiGaCr$_4$O$_8$ with $B_{\mathrm f}$ = 0.6 shows an indication 
of a second-order magnetic transition. However, the transition at $T_{\mathrm{N}}$ = 12.8~K 
is not realized, because the paramagnetic component completely disappears just before $T_{\mathrm{N}}$ 
as shown in Fig.~\ref{FIG3}(c). The observed transition is first order, and is probably an AF transition 
with lattice distortion. Indeed, the recent neutron scattering measurements indicate symmetry lowering of 
the lattice in the lower temperature phase~\cite{Nilsen}. 

\begin{figure}[tb]
	\includegraphics[width=0.6\linewidth]{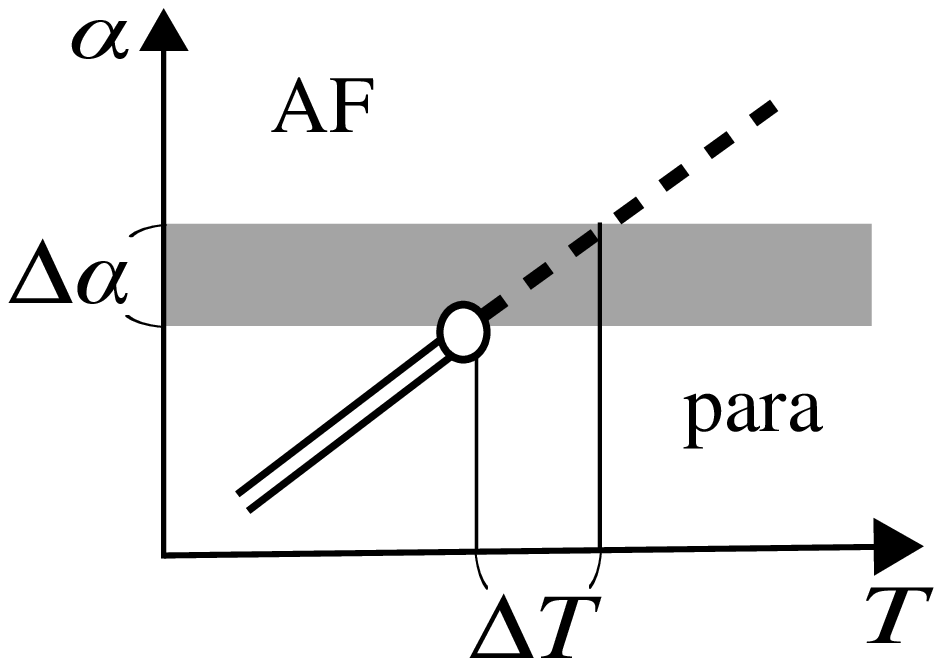}%
 \caption{Schematic phase diagram for LiGaCr$_4$O$_8$. On the vertical axis, $\alpha$ represents a
phenomenological tuning parameter of the Hamiltonian. The region $\Delta \alpha$ shows the
distribution of $\alpha$ in our sample. The dashed and double lines indicate first- and second-order
transitions, respectively, which are separated by the tricritical point indicated by the open circle.}
 \label{FIG4}  
 \end{figure}

Our results indicate that LiGaCr$_4$O$_8$ exhibits a first-order AF transition and at the same time, 
it is close to a critical point of a second-order transition. Such behavior can be most simply 
understood if the system is in the vicinity of a tricritical point, which separates a critical line into 
continuous and discontinuous regions in the phase diagram as schematically shown in 
Fig.~\ref{FIG4}. Here $\alpha$ represents a phenomenological tuning parameter of the system. 
For a microscopic spin Hamiltonian that describes the real material, $\alpha$ should be a combination 
of the parameters such as anisotropy, spin-lattice coupling, and multi-spin exchange. 
The distribution of the AF transition temperature 
($\Delta T$ in Fig.~\ref{FIG4}) observed in our experiments indicates a distribution of $\alpha$ ($\Delta \alpha$) 
in the sample as represented by the gray area in the phase diagram. 
The origin of the distribution $\Delta T$ or $\Delta \alpha$ is not understood yet. 
It could be due to extrinsic disorder in the sample. 

The susceptibility of the order parameter $\chi_m$ diverges at the tricritical point~\cite{Landau, Lawrie}, 
where  $1/T_1$ is expected to show a divergence due to critical slowing down. 
The distribution $\Delta \alpha$ for LiGaCr$_4$O$_8$ must be located on the first-order side. 
Even if $\alpha$ slightly deviates from the tricritical point to the first-order side, 
$\chi_m$ shows a divergence toward a slightly lower temperature than the first-order transition temperature~\cite{Lawrie}. 
Therefore, $1/T_1$ is expected to show divergence-like behavior just before the first-order transition. 
The coincidence between $T_{\mathrm{N}}$ and the temperature at which the paramagnetic 
component disappears can be described assuming that the tricritical point is 
located at the lower boundary of the gray area as shown in Fig.~\ref{FIG4}.
The location of the tricritical point just at the boundary of the distribution may not be a mere coincidence but has some physical 
mechanism.

Strong spin-lattice coupling in $A$Cr$_2$O$_4$ ({\textit A} = Zn, Cd) causes a first-order transition involving 
simultaneous AF spin order and lattice distortion at 12.5~K (ZnCr$_2$O$_4$) and 7.8~K (CdCr$_2$O$_4$)~\cite{Lee_Gasparovic}. 
The spin-Jahn-Teller effect has been proposed as the mechanism for these transitions.
However, the experimental information has been limited due to the first-order nature of these transitions. 
Our results indicate that LiGaCr$_4$O$_8$ also shows a simultaneous AF and structural phase transition due to strong spin-lattice coupling. 
Furthermore, the transition is very close to second-order unlike  $A$Cr$_2$O$_4$. 
If we can tune the control parameter $\alpha$, for example, by applying pressure or chemical doping 
and make the system cross the continuous critical line, 
it will allow us to study novel critical phenomena of a spin-lattice coupled transition in highly frustrated magnets.
To explore such possibility is a challenging future subject.

In summary, we performed $^{7}$Li-NMR measurements on  two 
breathing pyrochlore spin systems and showed that completely different magnetic 
properties are realized depending on the degree of breathing. 
In LiInCr$_4$O$_8$ with large breathing, spin gap behavior with $\varDelta$ = 31~K was observed in 1/$T_1$ 
at high temperature, which, however, is followed by a second-order AF transition at 13~K.
The AF transition is likely to be assisted by a structural transition at a slightly higher 
temperature that changes the symmetry of the lattice and releases frustration. In contrast, 
LiGaCr$_4$O$_8$ with smaller breathing does not show spin gap behavior but undergoes a 
first-order magnetic transition with a distribution in the transition temperature. 
A critical divergence of 1/$T_1$ in the coexisting paramagnetic phase indicates that 
LiGaCr$_4$O$_8$ is in the vicinity of a tricritical point.

\begin{acknowledgments}
We thank H. Tsunetsugu for stimulating discussion and useful comments on the interpretation of the 
experimental results. The work was supported by Grant-in-Aids for JSPS KAKENHI (B) (No. 25287083).
Y.~T. was supported by Japan Society for the Promotion of Science through Program for Leading Graduate Schools (MERIT).
\end{acknowledgments}


\renewcommand{\thefigure}{S\arabic{figure}}
\renewcommand{\theequation}{S\arabic{equation}}
\setcounter{figure}{0}
\setcounter{equation}{0}

\vspace{5mm}
\section*{SUPPLEMENTAL MATERIAL}
\subsection*{Line width of LiInCr$_4$O$_8$}

\begin{figure}[hb]
	\includegraphics[width=0.7\linewidth]{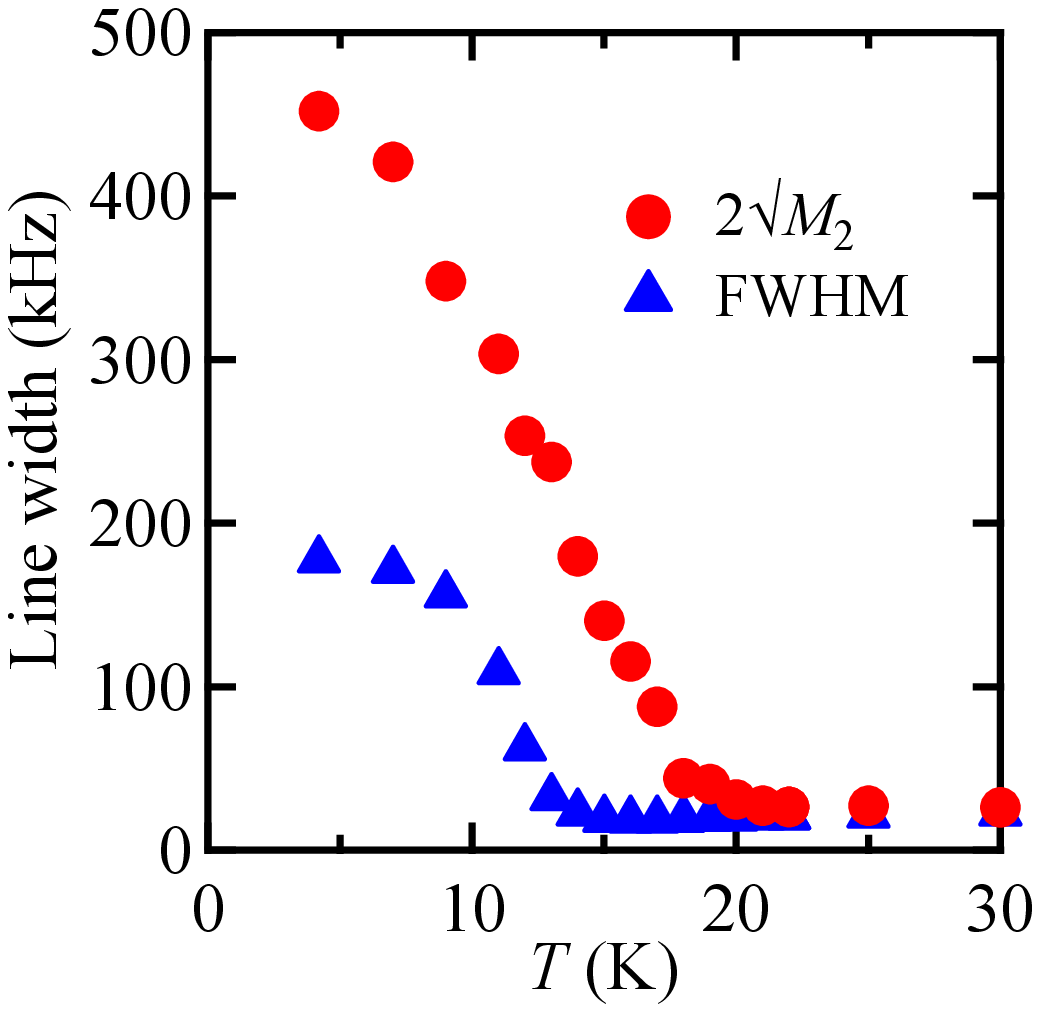}%
 \caption{(color online) Temperature dependence of the line width of NMR spectra of LiInCr$_4$O$_8$ at 2~T. 
 Blue triangles and red circles represent the FWHM and the $2\sqrt{M_2}$ measured with cooling, respectively.
}
 \label{FIGS1}  
 \end{figure}

By taking a closer examination of Fig.~\ref{FIG1} in the main text, we can see that the spectrum broadens in two steps with decreasing temperature.
First only the tails on both sides of the spectrum become broad with cooling while the main peak remains sharp down to 14~K. 
The width of the main peak then increase only below 14~K.
Figure~\ref{FIGS1} shows the temperature dependence of the full width half maximum (FWHM) and the square loot of the second moment $M_2$ obtained by
	\begin{equation}\label{eq:S1}
  	M_2=\int(f-f_0)^2I(f) \mathrm{d}f,
	\end{equation}
where $f$ is frequency,  $f_0$ is the center of gravity, and $I(f)$ is the NMR spectrum normalized as $\int I(f)\mathrm{d}f=1$.
With decreasing temperature, FWHM and $2\sqrt{M_2}$ start to grow near 14~K and 18~K, respectively.
$2\sqrt{M_2}$ is very sensitive to growth of the spectral tails.
This result suggests that a two-step transition occurs in LiInCr$_4$O$_8$, although we could not determine 
whether the increase of $2\sqrt{M_2}$ near 18~K arises from a structural transition or inhomogeneous spin freezing.

\end{document}